\documentclass[aps,prd,nofootinbib]{revtex4}
\usepackage{epsfig}
\usepackage{slashed}

\newcommand{\bel}[1]{\begin{equation}\label{#1}}
\newcommand{\bal}[1]{\begin{eqnarray}\label{#1}}

\newcommand{\be}{\begin{equation}}
\newcommand{\ee}{\end{equation}}
\newcommand{\ba}{\begin{eqnarray}}
\newcommand{\ea}{\end{eqnarray}}

\newcommand{\bes}{\begin{equation*}}
\newcommand{\ees}{\end{equation*}}

\begin{document}
\title{Deconfinement in the presence of a strong magnetic background: \\ 
an exercise within the MIT bag model}

\author{Eduardo S. Fraga\footnote{fraga@if.ufrj.br} and 
Let\'\i cia F. Palhares\footnote{leticia@if.ufrj.br}}

\affiliation{Instituto de F\'\i sica, Universidade Federal do Rio de Janeiro, \\
Caixa Postal 68528, Rio de Janeiro, RJ 21941-972, Brazil}

\begin{abstract}
We study the effect of a very strong homogeneous magnetic field $B$ on the thermal 
deconfinement transition within the simplest phenomenological approach: the MIT 
bag pressure for the quark-gluon plasma and a gas of pions for the hadronic sector. 
Even though the model is known to be crude in numerical precision and misses the 
correct nature of the (crossover) transition, it provides a simple setup for the discussion 
of some subtleties of vacuum and thermal contributions in each phase, and should 
provide a reasonable qualitative description of the critical temperature in the presence 
of $B$. We find that the critical temperature decreases, saturating for very large 
fields. 
\end{abstract}

\maketitle

\section{Introduction}

The area of in-medium strong interactions under extreme magnetic fields has gone 
through a fast development due to the combination of three features realized recently: 
experimental relevance, capability of being investigated on the lattice and a rich 
phenomenology.

Hot QCD matter exposed to sufficiently high magnetic fields is currently being generated 
in laboratory experiments. The setup of ultra-relativistic heavy ion collisions is such that 
extremely intense magnetic fields are produced in peripheral collisions, with a nonzero 
impact parameter. A beam of heavy ions accelerated to velocities very close to that of the 
light creates an enormous electric current. Non-central collisions induce, via a basic classical 
electrodynamics process, a maximum magnetic field at the center, exactly where the medium 
forms in such collisions. At RHIC the intensity of the magnetic fields created may reach 
$B \sim 10^{19}~$Gauss ($eB \sim6\,m_{\pi}^2$) \cite{magnetic-HIC}.

There have been many developments in the direction of using this extreme field to produce 
signatures to assess the properties of the new state of matter created in these experiments. 
For instance, the chiral magnetic effect \cite{CME}, which stimulated most of the recent 
development in the study of the phase diagram including an external magnetic field, takes 
advantage of the capability of magnetic fields for charge separation in space to render the 
possible formation of sphaleron-induced CP-odd domains in the medium measurable. 

From the theoretical point of view, the non-trivial role played by magnetic fields in the 
nature of phase transitions has been known for a long time \cite{landau-book}. Modifications 
in the vacuum of QED and QCD have also been investigated 
within different frameworks, mainly using effective 
models \cite{Klevansky:1989vi,Gusynin:1994xp,Babansky:1997zh,Klimenko:1998su,
Semenoff:1999xv,Goyal:1999ye,Hiller:2008eh,Rojas:2008sg}, 
especially the NJL model \cite{Klevansky:1992qe}, and chiral perturbation 
theory \cite{Shushpanov:1997sf,Agasian:1999sx,Cohen:2007bt}, but also resorting to the 
quark model \cite{Kabat:2002er} and certain limits of QCD \cite{Miransky:2002rp}. 
Interesting phases in dense systems \cite{Ferrer:2005vd,Son:2007ny}, as well as effects on 
the dynamical quark mass \cite{Klimenko:2008mg} were also considered. Nevertheless, the 
mapping of the new $T-eB$ phase diagram is still an open problem. Sufficiently large magnetic 
fields could allow for drastic phenomena, from shifting the chiral and the deconfinement phase 
transition lines 
\cite{Agasian:2008tb,Fraga:2008qn,Boomsma:2009yk,Fukushima:2010fe,Avancini:2011zz,Kashiwa:2011js,Chatterjee:2011ry,Skokov:2011ib} 
to transforming vacuum into a superconducting medium via $\rho$-meson 
condensation \cite{Chernodub:2010qx}. Most of the analyses so far were done within effective 
models, but several very interesting lattice QCD results are starting to 
appear \cite{lattice-maxim,lattice-delia,Braguta:2011hq,Bali:2011qj}.

In this paper we address the behavior of the critical temperature for deconfinement in the 
presence of a very large magnetic field within the simplest phenomenological approach: the 
MIT bag pressure for the quark-gluon plasma and a gas of pions for the hadronic 
sector. Even though the model is known to be crude in numerical precision and misses the 
correct nature of the (crossover) transition, it provides a simple setup for the discussion 
of some subtleties of vacuum and thermal contributions in each phase. As we will show, the 
procedure for subtracting purely magnetic contributions proves to be crucial. Moreover, if the influence 
of the magnetic field on the thermodynamics of both extreme energy domains is captured, it should 
furnish a reasonable qualitative description of the behavior of the critical temperature in the presence 
of $B$. Since we are not aiming at a precise description of the transition, but rather at 
the qualitative behavior of the critical temperature, $T_{c}$, as the magnetic field is increased, 
the simple model we adopt has the advantage of being very economic in terms of parameters 
to be fixed (essentially one) and other ingredients usually hard to control in more sophisticated 
effective models. Given the recent lattice results \cite{Bali:2011qj} that seem to indicate that 
$T_{c}$ goes down for increasing $B$, very differently from what has been obtained by 
effective model calculations and previous lattice simulations \cite{lattice-delia}, we believe 
this simple exercise within the MIT bag framework is valid. We find that the critical temperature 
decreases and saturates. 

The MIT bag model \cite{Chodos:1974pn} in a magnetic field was previously 
considered in Ref. \cite{Chakrabarty:1996te} to describe the stability and gross properties 
of quark matter in a different context. The pioneering treatment of deconfinement in the presence 
of a magnetic field presented in Ref. \cite{Agasian:2008tb} ends 
up essentially with the equivalent of a gas of pions plus a bag model description for the quark sector 
and computes the phase diagram. Nevertheless, our approach, as well as its predictions, differs 
qualitatively and quantitatively. More specifically, our renormalization procedure is very different, 
since we subtract purely $B$-dependent contributions to the pressure while Ref. \cite{Agasian:2008tb} 
does not. This difference has three crucial physical consequences: (i) our result is consistent with 
magnetic catalysis, while their result misses this feature. So, our result is consistent with those 
obtained in chiral perturbation theory, NJL and the linear sigma model; (ii) our critical temperature 
saturates for large $B$ whereas in their case it drops quadratically to zero, vanishing at $eB = 25 m_\pi^2$. 
To our knowledge, no lattice simulation or effective model approach has ever found a critical temperature 
that falls to zero, even for values of $eB$ much larger than that. The most recent lattice results feature a 
critical temperature that falls and then saturates for large $B$, a result that is in line with our findings and 
in stark contrast to the prediction of Ref. \cite{Agasian:2008tb}.

The paper is organized as follows. In Section II, we present a detailed computation of the renormalized 
pressures for both high- and low-energy sectors; in Section III, we show and discuss the $T-eB$ phase 
diagram; Section IV contains an explicit discussion on the subtleties of renormalization at finite $B$ 
and their relation to magnetic catalysis; Section V presents our conclusions.

\section{Renormalized pressures}

In the MIT bag model framework for the pressure of strong interactions, one has to evaluate 
the free quark pressure. The presence of a magnetic field in the $z$ direction, ${\bf B}=B\hat z$, 
affects this computation by modifying the dispersion relation to 
\begin{equation}
\omega_{\ell sf}(k_{z})=k_{z}^{2}+m_{f}^{2}+q_{f}B(2\ell+s+1)\equiv k_{z}^{2}+M_{\ell sf}^{2} \,,
\label{landau-levels}
\end{equation}
$\ell=0,1,2,\dots$ being the Landau level index, $s=\pm 1$ the spin projection, $f$ the flavor index, 
and $q_{f}$ the absolute value of the electric charge. In order to compute integrals over four-momenta and thermal sum-integrals one can use the following mappings, 
respectively \cite{Fraga:2008qn,Chakrabarty:1996te}:
\begin{eqnarray}
&&\int \frac{d^4k}{(2\pi)^4} \mapsto \frac{q_{f}B}{2\pi}\sum_{\ell=0}^\infty 
\int \frac{dk_0}{2\pi}\frac{dk_z}{2\pi} \, ,\\
T\sum_{n} &&\int \frac{d^3k}{(2\pi)^3} \mapsto \frac{q_{f}BT}{2\pi}\sum_n \sum_{\ell=0}^\infty 
\int \frac{dk_z}{2\pi} \, ,
\end{eqnarray}
where $\ell$ represents the different Landau levels and $n$ stands for the Matsubara 
frequency indices \cite{FTFT-books}. Since it has been shown that only very large magnetic 
fields can affect significantly the structure of the phase diagram for strong interactions 
\cite{Agasian:2008tb,Fraga:2008qn,Boomsma:2009yk,Fukushima:2010fe,lattice-delia,Bali:2011qj}, 
we compute the free quark pressure in the limit of very high magnetic fields, where it is also 
possible to simplify some analytic expressions. It is crucial to realize, however, that the lowest 
Landau level (LLL) approximation for the free gas pressure is not equivalent to the leading order 
of a large magnetic field expansion at all. For the zero-temperature, finite-$B$ contribution to the 
pressure, the LLL is the energy level which less contributes in the limit of large $B$; the result 
being dominated by high values of $\ell$. Nevertheless, the equivalence between the LLL 
approximation and the large $B$ limit remains valid for the temperature dependent part of the 
free pressure (as well as for the propagator), simplifying the numerical evaluation of thermal integrals.

This free magnetic contribution to the quark pressure has been considered in different contexts 
(usually, in effective field theories 
\cite{Fraga:2008qn,Boomsma:2009yk,Ebert:2003yk,Menezes:2008qt,Andersen:2011ip}) 
and computed from the direct knowledge of the energy levels of the system, 
Eq. (\ref{landau-levels}). The exact result, including all Landau levels, has to be computed from:
\begin{equation}
P_{q}=
2 N_c \sum_{\ell,s,f}\frac{q_f B}{2\pi}
\int \frac{dk_z}{2\pi} \bigg\{ 
\frac{\omega_{\ell sf}(k_z)}{2}
+T \ln\left[ 1+e^{-\omega_{\ell sf}(k_z)/T} \right]
\bigg\}
\, ,\label{bolhaquark}
\end{equation}
where the first term is a clearly divergent zero-point energy term and the other one is the 
finite-temperature contribution for vanishing chemical potential. Since $\omega_{\ell sf}$ 
grows with $B$, the largest the $\ell$ labeling the Landau level considered the larger the 
zero-point energy term becomes, being minimal for the LLL. Thus, in the limit of large $B$, 
the LLL approximation is inadequate. The decaying exponential dependence of the 
finite-temperature term on $\omega_{\ell sf}$, on the other hand, guarantees that the LLL 
dominates indeed this result for intense magnetic fields. 

To obtain a good approximation for the large $B$ limit of the free pressure, we choose to 
treat the full exact result and take the leading order of a $x_f\equiv m_f^2/2q_fB$ expansion in the 
final renormalized expression. Let us then discuss the treatment of the divergent zero-point 
term. Despite being a zero-temperature contribution, the first term in Eq. (\ref{bolhaquark}) 
can not be fully subtracted because it carries the modification to the pressure brought about 
by the magnetic dressing of the quarks. Using dimensional regularization and the zeta-function representation, which is also a type of regularization, for the sums over Landau levels and 
subtracting the pure vacuum term in $(3+1)$ dimensions, one arrives at:
\begin{equation}
P_{q}^{V}=
\frac{N_c}{2\pi^2} \sum_f (q_fB)^2\Big[
\zeta'\left(-1,x_f\right)
+\frac{1}{2}(x_f-x_f^2)\ln x_f+\frac{x_f^2}{4}-
\frac{1}{12}\big(
2/\epsilon+\ln (\Lambda^2/2q_fB)+1\big)
\Big] \, ,
\end{equation}
where a pole $\sim (q_fB)^2[2/\epsilon]$ still remains. This infinite contribution that survived the 
vacuum subtraction can be interpreted as a pure magnetic pressure coming from the artificial 
scenario adopted, with a constant and uniform $B$ field covering the whole universe (analogous 
to the case of a cosmological constant). In this vein, one may neglect all terms $\sim (q_f B)^2$ 
and independent of masses and other couplings (as done, e.g. in Refs. 
\cite{Fraga:2008qn,Menezes:2008qt,Andersen:2011ip}), concentrating on the modification of the pressure of the quark matter under investigation. This issue will be discussed in more detail in the last section of the paper.

The final exact result for the free pressure of magnetically dressed quarks is therefore 
\begin{equation}
\frac{P_{q}}{N_c}=
\sum_f \frac{(q_fB)^2}{2\pi^2}\Big[
\zeta'\left(-1,x_f\right)-\zeta'\left(-1,0\right)
+\frac{1}{2}(x_f-x_f^2)\ln x_f+\frac{x_f^2}{4}
\Big]
+T \sum_{\ell,s,f}\frac{q_fB}{2\pi^{2}}
\int dk_{z} 
\ln\left[ 1+e^{-\omega_{\ell sf}(k_z)/T} \right]
\,.
\end{equation}
In Refs. \cite{Fraga:2008qn,Menezes:2008qt,Andersen:2011ip}, the constant 
$\zeta'\left(-1,0\right)=-0.165421...$ was not subtracted. We believe, however, that it should also 
be excluded to be consistent with the subtraction of all terms that contribute to a (constant) 
pure magnetic pressure. In the case of pions, below, this procedure will ensure the 
realization of magnetic catalysis, i.e. an enhancement of chiral symmetry breaking, at zero 
temperature \cite{Gusynin:1994xp,Klimenko:1998su,Semenoff:1999xv,Miransky:2002rp}.
Clearly, if this term (of the form $\sim -\, \# \,(eB)^2$) is not subtracted, the pion contribution to the effective potential for the chiral condensate at large magnetic fields will eventually raise the minimum instead of lowering it.

In the limit of large magnetic field (i.e. $x_f=m_f^2/(2q_fB)\to 0$), we obtain
\begin{equation}
\frac{P_{q}}{N_c}\stackrel{{\rm large}~ B}{=}
\sum_f \frac{(q_fB)^2}{2\pi^2}\Big[
x_f\ln \sqrt{x_f}
\Big]
+T \sum_{f}\frac{q_fB}{2\pi^{2}}
\int dk_{z}
\ln\left[ 1+e^{-\sqrt{k_{z}^{2}+m_{f}^{2}}/T} \right]
\,.
\label{quark-pressure-largeB}
\end{equation}

Adding the free piece of the gluonic contribution and the bag constant ${\cal B}$, 
the pressure of the QGP sector in the presence of an intense magnetic field reads:
\begin{equation}
P^{B}_{\rm QGP}=
2(N_c^2-1)\frac{\pi^2T^4}{90}+P_{q}-{\cal B}
\,.
\end{equation}

\begin{figure}[htb]
\vspace{1cm}
\includegraphics[width=9cm]{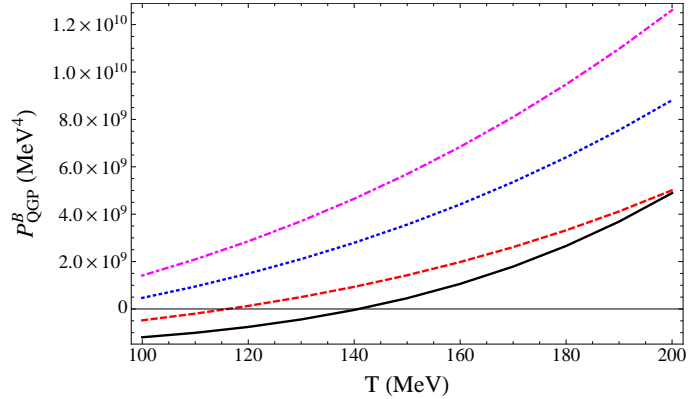}
\caption{Pressure in the QGP phase as a function of the temperature for different values 
of the magnetic field: $eB=0$ (black,solid,bottom), $20m_{\pi}^2,$ $40m_{\pi}^2,$ 
$60m_{\pi}^2$ (magenta,dash-dotted,top), where $m_{\pi}=138~$MeV 
is the vacuum pion mass.}
\label{bagquarkpressure}
\end{figure}

Figure \ref{bagquarkpressure} illustrates the behavior of $P_{\rm QGP}(T,B)$ for 
two flavors with $m_{u}=m_d=5~$MeV and $B^{1/4}=200~$MeV. 
Notice that, for $\sqrt{eB}$ much larger than 
all other energy scales, it is clear that the pressure in the QGP phase increases with 
the magnetic field. This seems to favor a steady drop in the critical temperature with 
increasing $B$ which would lead to a crossing of the critical line with the $T=0$ axis 
at some critical value for the magnetic field. However, the behavior of $T_{c}(B)$ 
also depends on how the pions react to $B$, and in this issue the treatment of the 
subtleties in the renormalization procedure as mentioned above is crucial. In fact, 
we will see that $T_{c}(B)$ goes down but tends to saturate, even for huge values of $B$.

In the confined sector, which we describe by a free pion gas, one may follow steps analogous 
to the ones described above in order to compute the contribution from the charged pions, 
which couple to the magnetic field. Again, using dimensional regularization and the zeta-function representation, and subtracting the pure vacuum term in $(3+1)$ dimensions, 
we arrive at
\begin{equation}
P_{\pi^{+}}+P_{\pi^{-}}=
-\frac{(e B)^2}{4\pi^2}\Big[
\zeta'\left(-1,\frac{1}{2}+x_{\pi}\right)-\zeta'\left(-1,\frac{1}{2}\right)
+\frac{x_{\pi}^{2}}{4} -x_{\pi}^{2} \ln \sqrt{x_{\pi}}
\Big]
-2 \frac{eB}{4\pi^{2}}
T \sum_{\ell} \int dk_{z} 
\ln\left[ 1-e^{-\sqrt{k_{z}^{2}+M^{2}_{\pi \ell}}/T} \right] \,,
\end{equation}
where $M^{2}_{\pi \ell}\equiv m_{\pi}^{2}+(2\ell +1)eB$ and $x_{\pi}\equiv m_{\pi}^{2}/(2eB)$. In 
this final expression all terms $\sim (q_f B)^2$ and independent of masses and other couplings 
were subtracted, as discussed before. 
Notice that the spin-zero nature of the pions guarantees that {\it all} charged pion modes in a magnetic field, differently from what happens with the 
quark modes, are $B$-dependent. So, in the large magnetic field limit the thermal integral 
associated with $\pi^{+}$ and $\pi^{-}$ is exponentially suppressed by an effective mass $\gtrsim (m_{\pi}^2+eB)$ and can be dropped. In this 
limit, we have
\begin{equation}
P_{\pi^{+}}+P_{\pi^{-}}\stackrel{{\rm large}~ B}{=}
-\frac{(e B)^2}{4\pi^2} \zeta^{(1,1)}(-1,1/2) ~x_{\pi} \,,
\label{pion-pressure-largeB}
\end{equation}
where $\zeta^{(1,1)}(-1,1/2)=-\ln(2)/2=-0.346574\cdots$. Neutral pions do not couple to the magnetic 
field and contribute only with a thermal integral of the form
\begin{equation}
P_{\pi^{0}}=
-\frac{T}{2\pi^{2}} \int dk k^{2} 
\ln\left[ 1-e^{-\sqrt{k^{2}+m_{\pi}^{2}}/T} \right] \,.
\end{equation}
\begin{figure}[htb]
\vspace{1cm}
\includegraphics[width=9cm]{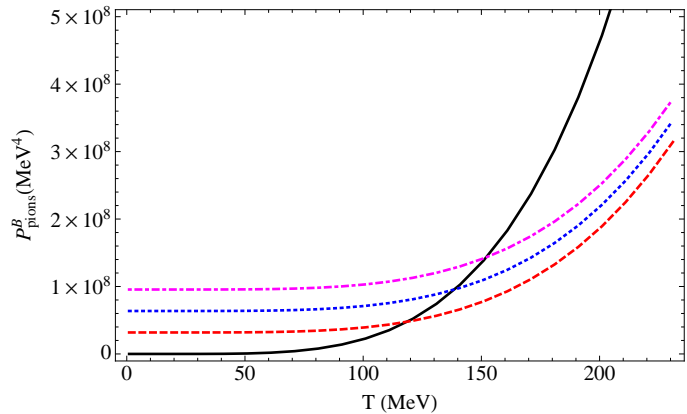}
\caption{Pressure in the pion gas phase as a function of the temperature for different values 
of the magnetic field: $eB=0$ (black,solid,bottom), $20m_{\pi}^2,$ $40m_{\pi}^2,$ 
$60m_{\pi}^2$ (magenta,dash-dotted,top), where $m_{\pi}=138~$MeV 
is the vacuum pion mass.}
\label{pionpressure}
\end{figure}

In Figure \ref{pionpressure} we show the pressure of the pion gas in the presence of 
a strong magnetic field. Again, for $\sqrt{eB}$ much larger than all other scales, the 
pressure rises with the magnetic field. This is a consequence of the subtraction of all 
terms that are independent of temperature, masses and other couplings in the 
renormalization process, which renders the pressure positive. As mentioned before, 
this is consistent with the phenomenon of magnetic catalysis at 
zero temperature. This plot presents, still, a couple of interesting 
features. First, differently from the quark case, the $B=0$ pressure takes over for 
temperatures of the order of the pion mass, which is not small and always enlarged 
by the presence of a magnetic field (given its scalar nature). Second, for large 
$T$, the magnetic pion pressures converge to $(1/3)$ of the $B=0$ pressure, since 
$\pi^{0}$ is the only degree of freedom that contributes thermally for large $B$.

\section{Phase diagram}

\begin{figure}[htb]
\vspace{1cm}
\includegraphics[width=9cm]{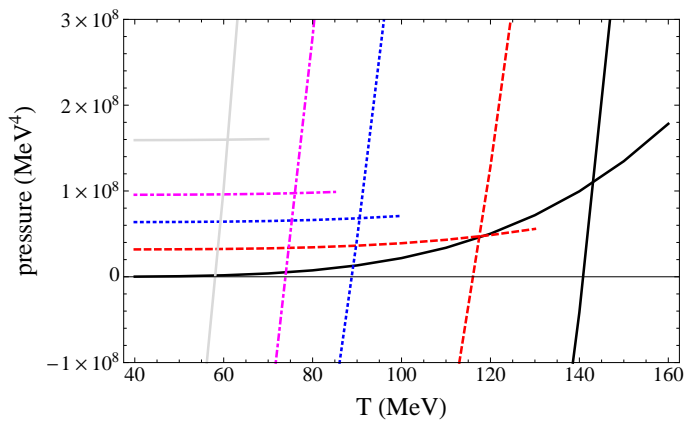}
\caption{Crossing pion gas and QGP pressures as functions 
of the temperature for different values 
of the magnetic field: $eB=0$ (black, solid, right-most), $20m_{\pi}^2,$ $40m_{\pi}^2,$ 
$60m_{\pi}^2$ (magenta,dash-dotted) and $eB=100m_{\pi}^2$ (gray, solid, left-most), 
where $m_{\pi}=138~$MeV is the vacuum pion mass.}
\label{crossingpressures}
\end{figure}

\begin{figure}[htb]
\vspace{1cm}
\includegraphics[width=9cm]{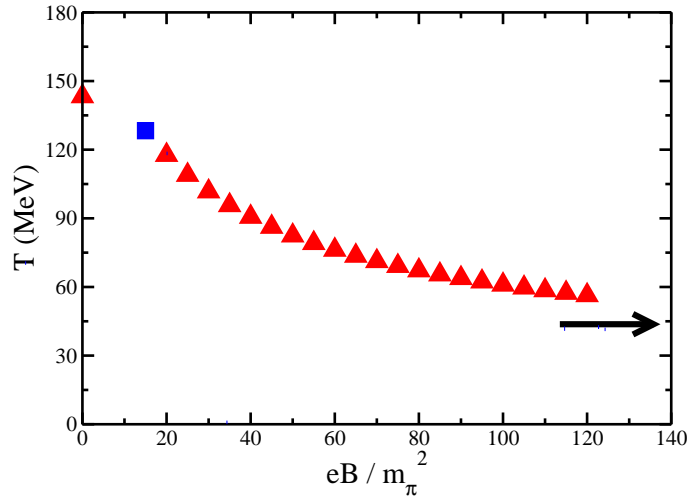}
\caption{Phase diagram in the presence of a strong magnetic field. We also keep the 
$T_{c}(B=0$) point. The blue square represents a very conservative estimate for 
the maximum value of $eB$ expected to be achieved in non-central collisions at the 
LHC with the formation of deconfined matter (it seems that much higher magnetic fields 
can be reached at the LHC due to the fluctuations in the distribution of protons inside the 
nuclei \cite{Bzdak:2011yy}, that being especially relevant to modifications of the 
vacuum \cite{Chernodub:2010qx}). 
The arrow marks the critical temperature for $eB\approx 210 m_{\pi}^2$ \cite{Vachaspati:1991nm}, 
expected to be found at the early universe.}
\label{phdiagramB}
\end{figure}

Each equilibrium phase should maximize the pressure, so that the critical line in 
the phase diagram can be constructed by directly extracting $T_{c}(B)$ from the equality 
of pressures. It is instructive, nevertheless, to consider a plot of the crossing pressures, 
as shown in Figure \ref{crossingpressures}. The figure shows, as expected, a decrease 
in the critical temperature (crossing points) as $B$ is increased due to the corresponding 
positive shift of the QGP pressure. However, $T_{c}$ seems to be saturating. One can 
see that the critical pressure (crossing point) goes down, 
but then it bends up again due to the increase in the pion pressure with $B$. This 
combination avoids a steady and rapid decrease of the critical temperature, as becomes 
clear in the phase diagram shown in Figure \ref{phdiagramB}. In fact, inspection of the 
zero-temperature limit of 
Eqs. (\ref{quark-pressure-largeB}) and (\ref{pion-pressure-largeB}) 
shows that there is no value of magnetic field that allows for a vanishing 
critical temperature.


The phase diagram in the plane $T-eB$ shows that the critical temperature for deconfinement 
falls as we increase the magnetic field. However, instead of falling with a rate that will 
bring it to zero at a given critical value of $eB$, it falls less and less rapidly, tending to 
saturate at large values of $B$. Interestingly, this qualitative behavior is very similar 
to the one obtained on the lattice in Ref. \cite{Bali:2011qj}\footnote{Of course, ignoring 
the caveat that our description necessarily predicts a first-order transition, as usual with 
the MIT bag model, and our numbers should be taken as rough estimates, as is always 
the case in effective models.}. 

Previous model descriptions \cite{Fraga:2008qn,Fukushima:2010fe,Skokov:2011ib}, have 
predicted either an increase or an essentially flat behavior for the deconfinement critical line 
as $B$ is increased to very large values, the same being true for previous lattice 
simulations \cite{lattice-delia}. Contrastingly, a significant decrease in the critical temperature 
as a function of $B$ was found in Ref. \cite{Agasian:2008tb}, where the authors build a model 
that at the end also behaves like a bag model. Their critical temperature, however, vanishes at a 
finite critical value of $eB_{c}\sim 25m_{\pi}^2$, featuring the disappearance of the confined phase 
at large magnetic fields. This phenomenon was not observed on the lattice (even for much larger 
fields) nor in any other effective model.  
A crucial difference in their approach is the subtraction procedure in which they keep the pure magnetic 
pressure terms $\sim(eB)^2$, in contradiction with the phenomenon of magnetic catalysis in the pion sector. 
As discussed in the next section, it can also bring unphysical different shifts in the free energies of 
the two phases.

Although our description of the deconfinement transition in the presence of an external 
large magnetic background is admittedly very simple, we believe it contains the relevant 
ingredients to provide a qualitative description of the behavior of the critical temperature 
with $B$. In fact, it seems that its focus on incorporating confinement (even 
if in the simplest fashion) makes it suited to describe the behavior of $T_{c}$ as a function 
of external parameters. Remarkably, we have previously achieved a successful description 
of the behavior of the critical temperature as a function of the pion mass and isospin chemical 
potential, as compared to lattice data, where chiral models failed even 
qualitatively \cite{Fraga:2008be,leticia-tese}. These pieces of evidence may indicate that the 
role played by confinement in guiding the functional behavior of $T_{c}$ could be central.

\section{Subtleties of renormalization at finite $B$ and magnetic catalysis}

In spite of the fact that the renormalization procedure in the presence of a constant and uniform magnetic field 
has proven to be very subtle and crucial for the phenomenological outcome for the phase structure, a more 
explicit discussion has usually been ommited in the literature. As is probably clear from the previous sections, 
since the subtractions performed in vacuum terms do not follow from the inclusion of the usual vacuum 
counterterms, they have to be physically guided. In what follows we present the physical arguments 
adopted here.

First of all, one notices that $B$-dependent $(1/\epsilon)$-divergences remain in the pressure even after 
the standard subtraction at $T=0$ and $B=0$. Therefore, these divergences have to be subtracted in 
an {\it ad hoc} fashion, which is always an unpleasant procedure in quantum field theory. To our 
knowledge, this is a common procedure in all effective approaches to this problem. As usual, such a 
subtraction allows for finite terms, which brings some ambiguity that must be resolved by physical 
constraints.

In this particular setting, we believe that these $B$-dependent, mass-independent  terms remain 
because one has forced a constant magnetic field everywhere and forever, a very unphysical setup, 
so an infinite purely $B$-dependent energy density (or purely magnetic pressure) lasts and has to be 
subtracted by hand. Pure vacuum ($B=0$) will never furnish an appropriate counterterm for that divergence. 
Physically, this could be interpreted as an indirect manifestation of the renormalization of the charges that 
generate $B$ (over which we have no control), since $B$ must be induced somehow by real 
currents. Those charges are not included in the general description, and the {\it ad hoc} subtraction seems 
to be the price one has to pay.

If we had a first-principle description, then the effect of a purely magnetic contribution to the pressure 
would only shift the effective potential as a whole. In particular, there would be no modification on 
relative positions and heights of different minima that represent different phases of matter. Thus, those 
terms should have no effect on phase transitions. This is also what happens when one computes 
directly an effective potential in models such as the linear sigma model, this terms being only a trivial 
shift\footnote{Trivial in the sense that it does not depend on the order parameter, so that it does not 
deform the potential. However, pressures (that correspond to the value of the effective potential at 
each minimum) are affected.}. In 
approaches that resort to different effective theories to compute the pressure in the two different 
phases, one has to subtract this purely magnetic contribution in each sector. Of course, a convenient 
choice of the renormalization scale ensures a common subtraction in the two phases.

We chose to subtract all purely magnetic terms in the pressures. This guarantees that the pion 
pressure grows with increasing magnetic field at zero temperature, which is consistent with the 
well-known phenomenon of magnetic catalysis. Furthermore, lattice simulations usually measure 
derivatives of the pressure with respect to temperature and quark mass, and do not access 
derivatives with respect to $B$. Therefore, purely $B$-dependent terms are not included in 
their results.

\section{Conclusions}

To summarize, although very simple in nature the MIT bag model description seems to contain 
the essential ingredients to describe the functional behavior of the critical temperature as observed 
in most recent first-principle lattice computations. Together 
with the fact that chiral models, even when coupled with the (static) Polyakov loop sector, seem 
to fail, this suggests that the critical temperature in QCD is a confinement-driven observable.

Of course, besides complementary effective model descriptions and lattice simulations, a 
first-principle perturbative QCD calculation, including strong interactions to some extent 
and the electromagnetic dressing fully, would provide more insight and a more solid ground 
to build a better understanding of the phase structure of strong interactions in the presence 
of an external magnetic field. We will report on results for the perturbative QCD pressure in 
a future publication \cite{magnetic-pressure}. Preliminary results and a full discussion of 
technical difficulties and subtleties introduced by the magnetic field in the finite-temperature 
loop calculation can be found in Ref. \cite{leticia-tese}.

\section*{Acknowledgments} 
We thank especially J.-P. Blaizot for very fruitful discussions on related matters. 
We also acknowledge M. N. Chernodub, M. D'Elia, G. Endrodi and X.-G. Huang for comments 
and discussions. 
This work was partially supported by CAPES, CNPq, FAPERJ and FUJB/UFRJ.


\end{document}